\documentclass{ws-mpla}
\usepackage[dvipsnames]{xcolor}
\usepackage{url}
\usepackage[super,compress]{cite}
\usepackage[verbose]{hyperref}
\hypersetup{colorlinks=false,allbordercolors=blue,pdfborderstyle={/S/U/W 1}}

\usepackage{hyperref}
\usepackage{cleveref}

\begin{document}

\thispagestyle{plain}

\title{Toward a Community Roadmap for High Energy Physics and \\ Artificial Intelligence in China and Beyond}

\markboth{WSPC}{Using World Scientific's style file}

\author{Tianji Cai\footnote{Corresponding author: \email{tianji\_cai@tongji.edu.cn}}}
\address{School of Physical Science and Engineering, Tongji University, Shanghai, China \\
State Key Laboratory of Autonomous Intelligent Unmanned Systems, MOE Frontiers Science Center for Intelligent Autonomous Systems, Tongji University, Shanghai, China}

\author{Ke Li}
\address{Institute of High Energy Physics, Chinese Academcy of Sciences, Beijing, China\\
University of Chinese Academy of Sciences, Chinese Academy of Sciences, Beijing, China}

\author{Teng Li}
\address{Institute of Frontier and Interdisciplinary Science, Shandong University, Shandong, China}

\maketitle

\begin{abstract}

Artificial Intelligence (AI) is rapidly transforming scientific research and has become central to many data-intensive disciplines. High Energy Physics (HEP), with its vast data volumes, complex theoretical structures, and precision-driven methodologies, lies at a particularly fertile intersection with modern AI. In this document, we present a community-informed overview of AI+HEP development in China and beyond, motivated in part by discussions at the 2025 Quantum Computing and Machine Learning Workshop in Qingdao, Shandong Province. We briefly review current AI activities across experimental, phenomenological, and theoretical HEP, along with key aspects of the research ecosystem. This work does not aim to represent the entire community, but rather reflects a partial and evolving snapshot informed by discussions and perspectives gathered from members of the broader AI+HEP community. We hope it serves as an initial roadmap to inform future coordinated efforts and to lay the groundwork for a more comprehensive community white paper.

\end{abstract}

\section{Introduction} \label{sec:intro}

Recent years have seen Artificial Intelligence (AI) emerging as a transformative force across science and technology, with ``AI for Science'' (AI4S) becoming a major focus of governmental and research initiatives worldwide~\cite{Genesis, DOEAI, EUAI, CNAI, RIKENAIP}. Substantial investments are now being directed toward integrating AI into scientific workflows, with the ambition to accelerate discovery and enable new modes of research. These global developments reflect a growing recognition of AI-driven science as a fundamental paradigm shift. At present, much of the attention within AI4S has been directed toward areas such as biology, medicine, materials science, and energy, where societal benefits are more direct. Fundamental physics, on the other hand, has received comparatively less emphasis in this broader narrative, and high energy physics (HEP) in particular remains a relatively specialized field with more limited visibility outside the scientific community. Nevertheless, a vibrant and growing international community is actively advancing research at the intersection of AI and HEP. This document aims to highlight selected developments and emerging needs within the AI+HEP landscape, with particular attention to ongoing efforts in China.

The connection between artificial intelligence and high energy physics is indeed long-standing, dating back to the late twentieth century when neural networks and boosted decision trees became integral components of collider data analysis and the standard experimental toolkit~\cite{Denby:1999kv, Roe:2004na}. This bond has expanded significantly with the advent of modern AI techniques, leading to a rapidly growing body of work at their intersection, encompassing both experimental and theoretical efforts~\cite{hepmllivingreview}. On the experimental side, modern collider and astroparticle observatories generate enormous volumes of complex data, making AI increasingly valuable for signal extraction, simulation, and inference, with applications ranging from event classification and particle identification to fast simulation. On the theoretical side, AI provides new tools for exploring complex mathematical structures, including scattering amplitudes, quantum field theory calculations, and lattice simulations. Conversely, HEP offers a uniquely structured and symmetry-rich environment for developing next-generation AI methodologies, particularly in areas such as interpretability, robustness, generalization, and the incorporation of first-principles constraints into learning systems.

These developments have long been supported by an active international community, predating recent large-scale government initiatives. Workshops and conference series, such as ML4Jets initiated in 2017~\cite{ML4Jets2017}, have played a central role in fostering collaboration, initially within collider physics and now across a broader range of HEP topics. Building on these bottom-up efforts, the field has increasingly attracted coordinated institutional support. In the United States, interdisciplinary research has been strengthened by dedicated institutes such as the the NSF AI Institute for Artificial Intelligence and Fundamental Interactions (IAIFI) launched in 2020~\cite{IAIFI}. Similar initiatives like the European Coalition for AI in Fundamental Physics (EuCAIF)~\cite{EuCAIF} and the Foundation of ``Machine Learning Physics'' in Japan~\cite{MLPhys} have promoted cross-disciplinary integration and investment in shared infrastructure and training. Together, these developments signal a transition of the field from primarily exploratory, community-driven research toward more systematic and coordinated developments of an AI-native scientific ecosystem.

Within China and the broader East Asian region, AI+HEP research has also experienced rapid growth, supported by an increasingly active and interconnected community. Volunteer-driven initiatives such as the ``AI+HEP in East Asia'' series~\cite{IHAPPIER} have established regional venues for exchange across particle physics, astrophysics, and cosmology, complemented by activities including regular journal clubs, joint seminars, and annual workshops. At the same time, many experimental collaborations and major institutions, including the Institute of High Energy Physics of the Chinese Academy of Sciences, have developed their own internal AI-focused working groups and seminar series to promote cross-disciplinary interaction. Despite the momentum, these efforts remain largely separated from each other, and a coherent, community-wide perspective has yet to be fully articulated. The 2025 Quantum Computing and Machine Learning Workshop in Qingdao~\cite{Qingdao} represents one such effort to bring the community together, providing a week-long forum for in-depth discussion that has motivated the present document as well as an ongoing community survey.

We aim to provide an overview of the current AI+HEP landscape, identify key research directions, and highlight priorities for future development, with an emphasis on the Chinese community. The remainder of this document is organized as follows. Section 2 reviews key research topics, covering applications in experiment, phenomenology, and theory, as well as emerging directions in general-purpose AI tools such as large language models and agent-based systems. Section 3 discusses the domestic community ecosystem, including research infrastructure and workforce development. Section 4 summarizes the main findings and outlines directions for future coordination and growth.

Along with the ongoing community survey, this work represents an initial step toward a more comprehensive white paper that will incorporate broader community input and inform future strategic planning. We emphasize that the perspectives presented here are selective and evolving, reflecting the views gathered by the authors and their collaborators rather than a complete representation of the field.

\section{Research Topics} \label{sec:topics}

For the purpose of the present discussion, we group AI+HEP research into four broad and overlapping categories. Experimental HEP emphasizes detector design, hardware systems, and experiment-specific data acquisition and processing pipelines. Phenomenology focuses on data analysis, simulation, and physics inference tasks shared across experiments, including emerging transferable frameworks such as foundation models. High energy theory explores AI applications to formal problems, as well as ways in which theoretical physics may inform new AI methodologies. Finally, general-purpose AI tools, including HEP-aware large language models and agent-based systems, may support research, training, and broader scientific workflows. The boundaries between these categories remain fluid, with substantial collaboration across subfields.

\subsection{AI in High Energy Experiment} \label{subsection:ex}

Modern high energy experiments (HEP-ex) generate datasets of unprecedented scale and complexity. Facilities such as the Large Hadron Collider at CERN, long-baseline neutrino observatories, cosmic-ray arrays, and dark-matter detectors produce petabyte-scale data streams in which rare physics signals are embedded within overwhelming backgrounds. China occupies a unique and rapidly expanding position in this global landscape. Several major experiments are currently in operation, including the Beijing Spectrometer Experiment (BESIII)~\cite{BESIII}, a high-luminosity $e^+e^-$ collider in the tau-charm region; the Jiangmen Underground Neutrino Observatory (JUNO)~\cite{JUNO}, designed to resolve the neutrino mass ordering; and the Large High Altitude Air Shower Observatory (LHAASO)~\cite{LHAASO}, a hybrid gamma-ray and cosmic-ray array. Next-generation facilities promise to further expand the scale and precision of future experiments, including proposed China-based projects such as the Super Tau-Charm Facility (STCF)~\cite{Peng:2020orp} and the Circular Electron Positron Collider (CEPC)~\cite{CEPC} as a potential Higgs factory.

Such experiments often produce high-dimensional, heterogeneous signals across millions of detector channels, including spatial coordinates, energy deposits, timing information, and waveform traces. These data form a complex multimodal landscape in which the relevant physical patterns are often subtle, distributed, and highly non-linear. Deep learning is particularly well suited to such settings, enabling the extraction of representations and correlations that are difficult to capture with traditional analysis strategies. As a result, machine learning has become deeply integrated into modern experimental workflows, from data acquisition and reconstruction to simulation and physics analysis.

This integration is already visible across a wide range of experimental tasks. Convolutional and graph neural networks are used for particle identification and track reconstruction in complex detector geometries, while generative models such as flow-based models accelerate detector simulation by orders of magnitude. Boosted decision trees and compact neural networks provide high-efficiency, low-latency decisions in hardware trigger systems, enabling intelligent data selection at the earliest stages of the workflow. Together, these applications have led to significant gains in efficiency and precision, allowing experiments to extract more information from increasingly complex data.

Looking ahead, AI is expected to play an even deeper role in real-time and hardware-level applications. The integration of machine learning models into trigger systems and edge-computing architectures could enable data-driven selection at the front end of experiments, helping preserve rare signals that might otherwise be lost. Anomaly detection algorithms deployed on FPGAs or GPUs close to detector readout may further provide continuous sensitivity to unexpected phenomena. At the same time, AI-driven approaches to detector design and optimization are beginning to emerge, enabling the co-design of hardware and algorithms within a unified framework.

China is well positioned to incorporate these developments into the design, construction, and operation of its planned experimental facilities. For next-generation projects such as STCF and CEPC, integrating AI methodologies at the design stage, rather than retrofitting them later, represents an important strategic opportunity. Co-optimizing detector geometry, readout electronics, and embedded AI algorithms from the outset could enable more intelligent real-time processing and improve the scientific reach of future experiments. Achieving this goal will require close collaboration among physicists, engineers, and AI researchers, as well as sustained investment in both hardware and software infrastructure.

\subsection{AI in High Energy Phenomenology} \label{subsection:ph}

High energy phenomenology (HEP-ph) plays an important role in the AI+HEP landscape by connecting experimental data with theoretical interpretation. Many phenomenological tasks, including event classification, reconstruction, unfolding, parameter inference, and simulation, share common structures across collider, neutrino, astroparticle physics, and cosmology. This broad applicability, together with the data-rich nature of many phenomenological problems, has made HEP-ph both an early adopter of machine learning and an active area for its continued development and application.

Progress in HEP-ph has been driven largely by task-specific machine learning models tailored to well-defined physics problems. Classification tasks, especially collider jet tagging~\cite{favaro_2025_10878355}, have served as an early and influential testing ground for modern AI methods in HEP, evolving from high-level observables to increasingly expressive particle-level representations. More broadly, targeted approaches encompass a wide range of applications, including generative models for fast simulation and density estimation, simulation-based inference techniques for parameter extraction, and physics-informed architectures that incorporate symmetries and theoretical constraints. These methods have improved both computational efficiency and analysis sensitivity, but their task-specific nature also highlights a broader challenge: methodologies often remain fragmented across analyses, with limited transferability between problems and experiments.

This bottleneck motivates interest in more general and reusable models. Recent work has begun to explore foundation models for particle physics, in which a single model is pre-trained on large collections of experimental and simulated data and adapted to a wide range of  downstream tasks such as classification, fast simulation, trigger emulation, anomaly detection, and detector calibration~\cite{PhysRevD.111.L051504, Birk_2024}. Early studies suggest that such models can learn shared representations of physical systems, reduce the need for repeated task-specific training, and enable transfer across tasks or experiments. In the long term, these models may serve as reusable tools for HEP analysis, analogous in spirit to large language models in language-related domains~\cite{Barman_2025}. Realizing this vision will require coordinated efforts in data standardization, benchmarking, large-scale model development, and careful model verification, as well as close interaction with experimental collaborations.

Several directions are likely to shape the next phase of progress. These include advances in representation learning aimed at uncovering physically meaningful latent structures; improved interpretability and uncertainty quantification, which are essential for connecting machine learning outputs to theoretical understanding; and the integration of AI into end-to-end analysis pipelines, where simulation, inference, and parameter estimation are treated within a unified framework. In this context, emerging paradigms such as agent-based systems may further enable the automation of complex analysis workflows, a direction discussed in more detail in Section 2.4. Together, these developments point toward a transition from isolated machine learning applications to a more unified paradigm in which AI becomes an integral component of the scientific reasoning process.

\subsection{AI in High Energy Theory} \label{subsection:th}

The application of AI to high energy theory (HEP-th) extends machine learning beyond the familiar ``big data'' paradigm. Unlike largely data-driven applications in phenomenology, HEP-th focuses on structured mathematical objects, symbolic expressions, and formal reasoning tasks governed by exact constraints such as symmetries, conservation laws, and consistency conditions. These features define a distinct regime for AI, where the central challenge is less pattern recognition in large datasets than navigating structured spaces of mathematical possibilities. In this context, AI is not only a tool for computation, but increasingly a potential assistant for formal reasoning itself.

A broad class of theoretical problems can be framed as search problems in vast, highly constrained abstract spaces. Examples include the exploration of scattering amplitudes, the evaluation of Feynman integrals, the construction of effective field theories, and the study of lattice field theory configurations. These problems often involve intricate algebraic relations, large combinatorial structures, and complex optimization landscapes where traditional analytic or numerical approaches can become intractable. They are therefore closely connected to areas of AI concerned with optimization, combinatorial search, and structured decision-making, with machine learning serving as a tool for guided exploration rather than purely statistical modeling.

Current efforts in this direction are diverse and largely exploratory. In particular, reinforcement learning (RL) is emerging as a promising paradigm for tackling structured theoretical problems, where a neural-network-based agent iteratively explores constrained solution spaces guided by reward signals derived from mathematical or physical consistency. More recently, there has been growing interest in integrating large language model (LLM)-based agents into such frameworks, where the agent proposes candidate solutions, evaluates them against first-principles criteria, and refines its strategy through iterative feedback. These RL- and agent-based approaches provide new ways to efficiently navigate complex search spaces, with the potential to uncover hidden structures or simplified representations that are difficult to obtain through conventional techniques.

The interaction between AI and high energy theory is bidirectional, with the stringent demands of theoretical physics beginning to shape the development of AI methodologies themselves. The emphasis on exact reasoning, symbolic manipulation, and structured knowledge has motivated growing interest in neuro-symbolic methods, hybrid approaches that combine neural networks with symbolic reasoning, and learning paradigms that incorporate explicit constraints. Recent developments, including early prototypes of AI-assisted theorem discovery and domain-specific agent systems for theoretical physics, suggest a growing convergence between AI research and the needs of the HEP-th community.

A particularly important direction within this trend is the development of general AI systems with enhanced reasoning capabilities. Compared to applications in data analysis, where predictive performance is often the primary objective, theoretical applications require models that produce interpretable, verifiable, and logically consistent outputs. This has led to increasing exploration of AI and agent systems designed for structured reasoning and step-by-step inference. In this context, emerging frameworks such as AI-assisted mathematical discovery systems and HEP-specific agent architectures represent early steps toward integrating AI more deeply into the process of theoretical research.

This emerging landscape offers important opportunities. By combining theoretical expertise with new AI methodologies, the HEP-th community may contribute not only to applications of AI, but also to its conceptual and methodological development. More broadly, the interaction between AI and high energy theory may support new modes of scientific computing, where machine learning serves not only to accelerate calculations, but also to support reasoning, abstraction, and the generation of new theoretical insights.

\subsection{General AI Tools for HEP: LLMs and Agents} \label{subsection:tools}

The recent development of large language models (LLMs) and agent-based AI systems has opened new possibilities for building general-purpose tools for high energy physics. Unlike task-specific machine learning models, these systems can integrate multiple capabilities, including natural language understanding, code generation, symbolic reasoning, and workflow automation. As a result, they may become useful across experiment, phenomenology, and theory, as well as in training and education.

One natural direction is the development of HEP-aware language models adapted to the technical language, literature, software ecosystems, and data formats of high energy physics. Such models could assist with paper reading, documentation, code development, and communication across subfields. In the longer term, domain-specific LLMs may serve as interfaces between researchers and complex computational tools, helping users navigate simulation frameworks, detector and analysis software, databases, and theoretical calculations. Their value will depend not only on model scale, but also on the quality of domain-specific data, the reliability of retrieved information, and the ability to produce verifiable, reproducible, and scientifically meaningful outputs.

A second and more ambitious direction is the design of AI agent systems for scientific workflows. Such systems combine LLMs with external tools, memory, planning, execution, and feedback mechanisms, enabling multi-step tasks under human supervision. They could support general research activities, such as literature review, hypothesis generation, and writing assistance, as well as HEP-specific workflows, including detector monitoring, simulation setup, data selection, systematic studies, and result interpretation. Early prototypes, such as Dr.\ Sai~\cite{he2026drsaiagenticairealworld}, Aether~\cite{Aether}, and other domain-oriented agent frameworks, illustrate the growing interest in using agentic AI to automate parts of experimental and theoretical research pipelines.

The appeal of agentic systems is particularly strong because many research workflows are both highly structured and technically demanding~\cite{hepscript}. Experimental analyses often follow well-defined procedures, yet require substantial effort in data handling, validation, uncertainty estimation, and reproducibility checks. Theoretical and phenomenological studies likewise involve repeated cycles of calculation, simulation, fitting, and interpretation. In these settings, AI agents are best viewed not as replacements for researchers, but as tools for handling repetitive or technically detailed tasks, allowing scientists to focus more on physical interpretation, model building, and conceptual reasoning~\cite{opinionpaper}.

A future vision for agentic AI is to combine it with deep learning for a more global and systematic exploration of experimental data. Rather than applying individual models to isolated analysis tasks, future agent systems could coordinate suites of specialized subagents across many final states, control regions, signal hypotheses, and parameter spaces. Such systems could help organize large-scale scans of accessible physics processes, re-examine Standard Model channels, and search for subtle deviations that might otherwise be missed in narrower analyses. While human judgment would remain essential, these tools could accelerate the discovery cycle and extend the physics reach of large experimental programs.

At the same time, the development of LLM- and agent-based tools raises important challenges. Reliability, hallucination control, reproducibility, interpretability, data privacy, and integration with existing software frameworks are all essential concerns. For scientific applications, it is not sufficient for an AI system to produce plausible answers; its outputs must be traceable, testable, and consistent with domain knowledge. Progress will therefore require close collaboration between AI researchers and HEP domain experts, as well as benchmarks and evaluation protocols tailored to scientific reasoning, code correctness, numerical accuracy, and workflow reliability.

General AI tools may provide a bridge between the specialized applications discussed in the previous subsections and a more integrated AI-native research environment. Advancing this direction will require curated domain datasets, standardized software interfaces, scalable computing resources, and open evaluation platforms. If developed responsibly, HEP-aware LLMs and agent systems could help lower technical barriers, improve reproducibility, automate routine workflows, and support systematic exploration across parameter spaces, thus enhancing both the efficiency and scientific reach of high energy physics.

\section{Community Ecosystem} \label{sec:community}

The sustainable development of AI+HEP depends critically on an inclusive, collaborative, and well-supported community ecosystem, built around shared infrastructure, standardized data and software frameworks, and organizational structures for interdisciplinary collaboration. These topics formed an important part of the 2025 Qingdao workshop discussions, and this section summarizes the main themes that emerged.

\subsection{Research Infrastructure} \label{subsection:infra}

One critical factor is the availability of robust and accessible infrastructure. While the international community has made substantial progress in developing open datasets, shared benchmarks, and common software tools, the corresponding ecosystem in China is at an earlier stage of consolidation. Key challenges include data accessibility and computing resources. Many datasets are not yet available in standardized, machine-learning-ready formats, limiting their broader utility, while access to high-performance computing resources remains uneven across institutions. These constraints create barriers to entry, particularly for smaller groups and early-career researchers, underscoring the need for more coordinated infrastructure development.

Three specific concerns were repeatedly raised during the workshop discussions. The first is the need for shared data repositories, standardized benchmarks, and common software frameworks that can support collaboration across experiments and institutions. The second concerns sustained investment in computing infrastructure, from centralized facilities to distributed cloud-based systems, to support large-scale machine learning workloads. Finally, participants emphasized the growing need to account for environmental impact and energy efficiency as computational requirements continue to increase.

\subsubsection{Open Data}

The workshop discussions highlighted open data as a central infrastructure priority. A majority of participants rated open data as ``very important'' or ``extremely important'' for both current and future research, while emphasizing that the issue extends beyond data availability alone. The need for standardized formats, well-documented metadata, curated benchmarks, and reproducible workflows was emphasized. As an initial proposal, a unified open-data framework for AI+HEP research in China was discussed, with the following components:
\begin{itemize}
    \item Dual-scale datasets, including trial-sized samples suitable for small workstations for method development and educational purposes, as well as full-scale datasets hosted on cloud or high-performance platforms for large-model training.
    \item Standardized data formats and metadata schemas designed to facilitate interoperability across experiments.
    \item Curated benchmark datasets and tasks that enable fair comparison of AI methods.
    \item Integrated analysis workflows and reproducible pipelines accompanying each dataset.
\end{itemize}

\subsubsection{Computing Infrastructure}

Computational resources were identified as one of the most pressing bottlenecks. The rise of larger models, agentic AI workflows, and foundation-model-like approaches is expected to place increasing demands on GPU- and accelerator-based computing. This motivates calls for early planning and coordinated investment in computing infrastructure, including:
\begin{itemize}
    \item Shared computing platforms dedicated to AI+HEP research, hosted by central institutions but accessible to researchers nationwide.
    \item A federated computing network connecting existing institutional clusters, allowing unused resources to be shared across institutions.
    \item Strategic partnerships with cloud service providers to support subsidized access to cloud computing for academic use.
    \item Standardized containerized environments and workflow-management tools to ensure reproducibility across different computing platforms.
\end{itemize}

\subsubsection{Environmental Impact and Green AI}

A less frequently discussed but increasingly relevant issue is the environmental impact of large-scale AI computing for fundamental research. As computational demands grow, ``green AI'' principles---energy efficiency, carbon-footprint reduction, and sustainable computing practices---are becoming more relevant to the HEP community. Given the limited quantitative estimates of the environmental cost of AI+HEP research, more systematic study and transparent reporting were emphasized. Possible directions include:
\begin{itemize}
    \item Developing standardized methods for measuring and reporting the energy use and carbon footprint of AI models and computing workflows.
    \item Sharing trained models, datasets, and workflows to reduce unnecessary duplication of large-scale training.
    \item Encouraging efficient model architectures, training strategies, and resource-allocation practices.
    \item Promoting the use of renewable-energy-powered computing facilities where available.  
\end{itemize}

\subsection{Workforce Development} \label{subsection:workforce}

The long-term development of AI+HEP requires a workforce that can move effectively between physics questions and AI methodologies. Although cross-disciplinary literacy is growing on both sides, important gaps remain in translating between domain-specific physics needs and modern AI practice. Addressing these gaps will require sustained efforts in collaboration, organizational support, and talent development.

The workshop discussions highlighted institutional boundaries and mismatched incentive structures as recurring barriers to effective cross-disciplinary collaboration. Participants suggested that more formal joint research programs between physics and other departments, such as computer science, could help address this issue, especially through dual mentorship, co-designed projects, and shared evaluation criteria that value both scientific insight and AI methodological innovation. Partnerships with industry were viewed as valuable, particularly for expertise in large-scale machine learning systems, MLOps, and cloud computing, while HEP can in turn offer scientifically rigorous benchmark problems and high-impact research applications.

On the education side, the need for more structured AI+HEP training pathways was emphasized. While introductory machine learning courses are increasingly available, field-specific skills are still often acquired through self-study, informal mentoring, or project-based experience. A more sustainable talent pipeline could be supported by interdisciplinary curricula combining core physics, scientific computing, and modern AI methods, potentially including undergraduate minors, graduate specializations, and dedicated short courses or summer schools. For established researchers transitioning into AI+HEP, intensive schools and coding bootcamps were viewed as useful ways to lower entry barriers. Hackathons, research internships, and public lectures were also highlighted as mechanisms for attracting young students and building broader awareness of AI+HEP research.

\section{Outlook and Conclusion} \label{sec:outlook}

Artificial Intelligence is now an integral component of high energy physics, influencing all stages of the scientific workflow from experimental design and data acquisition to theoretical modeling and interpretation. What began as a collection of early, task-specific applications has evolved into a rapidly expanding and multifaceted research area, supported by both bottom-up community efforts and more recent institutional initiatives. These developments reflect a growing integration of AI into scientific practice and an increasing recognition of its broader methodological implications.

The AI+HEP community in China and the broader East Asian region is at a particularly dynamic stage of development, with its strong and unique experimental programs and a rapidly growing AI ecosystem in both academia and industry. Realizing this potential will require strategic coordination, sustained investment in infrastructure, and a commitment to global interdisciplinary collaboration. This motivates the need for structured efforts to better understand the current state of the field, identify common challenges, and articulate a shared vision.

As an initial step toward this goal, a community survey was launched in November 2025~\cite{SurveyA, SurveyB}. The survey is designed to collect input from researchers across different subfields, career stages, and institutional backgrounds, with the aim of building a more comprehensive and representative picture of AI+HEP activities and needs. At the time of writing, the survey has received more than 30 responses and remains open for further contributions.

The survey consists of two complementary parts, combining multiple-choice and open-ended questions. Part A covers essential and mandatory questions designed to capture a broad overview of the community and reflects the main themes discussed in this document. It requires less than 30 minutes to complete and is organized into several sections, each targeting a core aspect of the current research landscape:
\begin{itemize}
    \item \textbf{Background and research profile}, including demographics, institutional and geographic distribution, primary research areas, and level of involvement in AI+HEP research.
    \item \textbf{Current research activities and technical focus}, including both physics-driven topics (e.g., experimental analyses, phenomenology, and theory) and AI methodology (e.g., model classes, problem types, and application domains).
    \item \textbf{Assessment of AI capabilities and limitations}, including perceived effectiveness of existing AI approaches for physics applications, and gaps between current methods and scientific needs.
    \item \textbf{Challenges and resource constraints}, including limitations related to expertise, manpower, computing resources, data access, collaboration opportunities, and funding.
    \item \textbf{Community and collaboration}, including existing modes of interaction, participation in workshops and networks, and opportunities for broader collaboration.
    \item \textbf{Infrastructure and ecosystem needs}, including requirements for computing platforms, shared tools, and data resources, as well as the role of coordinated infrastructure.
    \item \textbf{Education and training}, including availability and quality of interdisciplinary training, and the importance of developing AI-related skills within the HEP community.
    \item \textbf{Future outlook and strategic priorities}, including expectations for the evolution of AI+HEP, both globally and within China, and perspectives on long-term development.
\end{itemize}

Part B is optional and consists of open-ended questions that complement the structured overview in Part A. It invites more detailed qualitative input on research directions, collaboration models, infrastructure design, funding mechanisms, education, and international engagement. By organizing the survey along these dimensions, we aim to translate the discussions presented in this roadmap into a more systematic understanding of the field, capturing not only current research activities but also structural bottlenecks, emerging opportunities, and possible directions for coordination beyond the level of individual projects.

Looking ahead, this roadmap and the ongoing survey represent an initial step toward a broader, community-driven effort. We welcome further participation in the survey, with the goal of exceeding 100 responses to support more statistically meaningful insights. Future work will synthesize the survey results, expand community input, and refine research and infrastructure priorities. These results will be presented in a subsequent white paper aimed at reflecting a broader consensus and providing concrete guidance for coordinated AI+HEP initiatives.

Given the inherently global nature of high energy physics, progress in AI+HEP depends on both local development and active international collaboration and open scientific exchange. The efforts described here should therefore be viewed as part of a broader cooperative endeavor to integrate AI into fundamental physics in ways that advance both fields. This co-evolution may not only enhance our ability to analyze large-volume data, but also reshape how scientific knowledge is generated. As a field grounded in open collaboration and the pursuit of fundamental understanding, HEP is well positioned to help foster a research culture in which advances in AI are guided by transparency, rigor, long-term scientific value, and broad societal benefit.

\section*{Acknowledgements}

The authors would like to thank Cheng-Wei Chiang, Xingtao Huang, Liang Li, Tao Liu, Yanqing Ma, and Manqi Ruan for their valuable feedback on the survey design. We are also grateful to all survey participants for their contributions, and to the attendees of the 2025 Qingdao workshop for their engagement in the discussions. The authors acknowledge the use of ChatGPT (OpenAI) for assistance with language editing and improving the clarity of the manuscript. All scientific ideas and content are the responsibility of the authors. The work of TC is supported by the Fundamental Research Funds for the Central Universities and sponsored by Shanghai Pujiang Programme. The work of Ke Li is supported by the Strategic Priority Research Program of Chinese Academy of Sciences under Grant XDA0480600. The work of Teng Li is supported by the National Natural Science Foundation of China (Grant No. 12341504, No. 12025502, No. 12475198 and No. 12175124).

\bibliographystyle{ws-mpla}
\bibliography{main}

@misc{Genesis,
  Author = "{DOE Genesis Mission}",
  howpublished = {\url{https://www.energy.gov/undersecretaryforscience/genesis-mission/genesis-mission?utm_source=chatgpt.com}},
  note = {Accessed: 2026-05-01}
}

@misc{DOEAI,
  Author = "{DOE Office of Science AI Initiative}",
  howpublished = {\url{https://science.osti.gov/Initiatives/AI?utm_source=chatgpt.com}},
  note = {Accessed: 2026-05-01}
}

@misc{EUAI,
  Author = "{European AI in Science Strategy}",
  howpublished = {\url{https://research-and-innovation.ec.europa.eu/strategy/strategy-research-and-innovation/our-digital-future/european-ai-science-strategy_en}},
  note = {Accessed: 2026-05-01}
}

@misc{CNAI,
  Author = "{China launches project to promote use of AI in sci-tech research}",
  howpublished = {\url{https://stcsm.sh.gov.cn/eng-news/20230414/35025c43a6f34ffba7ce2bf0c45c4822.html}},
  note = {Accessed: 2026-05-01}
}

@misc{RIKENAIP,
  Author = "{RIKEN Center for Advanced Intelligence Project}",
  howpublished = {\url{https://aip.riken.jp}},
  note = {Accessed: 2026-05-01}
}

@article{Denby:1999kv,
    author = "Denby, Bruce H.",
    title = "{Neural networks in high-energy physics: A ten year perspective}",
    doi = "10.1016/S0010-4655(98)00199-4",
    journal = "Comput. Phys. Commun.",
    volume = "119",
    pages = "219--231",
    year = "1999"
}

@article{Roe:2004na,
    author = "Roe, Byron P. and Yang, Hai-Jun and Zhu, Ji and Liu, Yong and Stancu, Ion and McGregor, Gordon",
    title = "{Boosted decision trees, an alternative to artificial neural networks}",
    eprint = "physics/0408124",
    archivePrefix = "arXiv",
    doi = "10.1016/j.nima.2004.12.018",
    journal = "Nucl. Instrum. Meth. A",
    volume = "543",
    number = "2-3",
    pages = "577--584",
    year = "2005"
}

@misc{hepmllivingreview,
    Author = "{HEP ML Community}",
    title = "{A Living Review of Machine Learning for Particle Physics}",
    url={https://iml-wg.github.io/HEPML-LivingReview/},
    note = {Accessed: 2026-05-01}
}

@misc{ML4Jets2017,
  Author = "{Machine Learning for Jet Physics (2017)}",
  howpublished = {\url{https://indico.physics.lbl.gov/event/546/}},
  note = {Accessed: 2026-05-01}
}

@misc{IAIFI,
    Author = "{The NSF AI Institute for Artificial Intelligence and Fundamental Interactions (IAIFI)}",
    howpublished = {\url{https://iaifi.org}},
    note = {Accessed: 2026-05-01}
}

@misc{EuCAIF,
    Author = "{European Coalition for AI in Fundamental Physics (EuCAIF)}",
    howpublished = {\url{https://eucaif.org}},
    note = {Accessed: 2026-05-01}
}

@misc{MLPhys,
    Author = "{Foundation of ``Machine Learning Physics'' (MLPhys)}",
    howpublished = {\url{https://mlphys.scphys.kyoto-u.ac.jp/en/}},
    note = {Accessed: 2026-05-01}
}

@misc{IHAPPIER,
    Author = "{AI+HEP in East Asia}",
    howpublished = {\url{https://ai-hep.github.io}},
    note = {Accessed: 2026-05-01}
}

@misc{Qingdao,
    Author = "{Quantum Computing and Machine Learning Workshop 2025}",
    howpublished = {\url{https://indico.ihep.ac.cn/event/25857/}},
    note = {Accessed: 2026-05-01}
}

@misc{BESIII,
    Author = "{Beijing Spectrometer (BESIII) Experiment}",
    howpublished = {\url{http://bes3.ihep.ac.cn}},
    note = {Accessed: 2026-05-01}
}

@misc{JUNO,
    Author = "{Jiangmen Underground Neutrino Observatory}",
    howpublished = {\url{https://juno.ihep.cas.cn}},
    note = {Accessed: 2026-05-01}
}

@misc{LHAASO,
    Author = "{Large High Altitude Air Shower Observatory}",
    howpublished = {\url{https://english.ihep.cas.cn/lhaaso/}},
    note = {Accessed: 2026-05-01}
}

@article{Peng:2020orp,
    author = "Peng, Hai Ping and Zheng, Yang Heng and Zhou, Xiao Rong",
    title = "{Super Tau-Charm Facility of China}",
    doi = "10.7693/wl20200803",
    journal = "Physics",
    volume = "49",
    number = "8",
    pages = "513--524",
    year = "2020"
}

@misc{CEPC,
    Author = "{Circular Electron Positron Collider}",
    howpublished = {\url{http://cepc.ihep.ac.cn}},
    note = {Accessed: 2026-05-01}
}

@dataset{favaro_2025_10878355,
  author       = {Favaro, Luigi and
                  Plehn, Tilman and
                  Kasieczka, Gregor},
  title        = {TopTagXL: A top quark tagging dataset for large
                   scale studies in jet tagging
                  },
  month        = aug,
  year         = 2025,
  publisher    = {Zenodo},
  version      = {1.0.0},
  doi          = {10.5281/zenodo.10878355},
  url          = {https://doi.org/10.5281/zenodo.10878355},
}

@article{PhysRevD.111.L051504,
  title = {Solving key challenges in collider physics with foundation models},
  author = {Mikuni, Vinicius and Nachman, Benjamin},
  journal = {Phys. Rev. D},
  volume = {111},
  issue = {5},
  pages = {L051504},
  numpages = {6},
  year = {2025},
  month = {Mar},
  publisher = {American Physical Society},
  doi = {10.1103/PhysRevD.111.L051504},
  url = {https://link.aps.org/doi/10.1103/PhysRevD.111.L051504}
}

@article{Birk_2024,
doi = {10.1088/2632-2153/ad66ad},
url = {https://doi.org/10.1088/2632-2153/ad66ad},
year = {2024},
month = {aug},
publisher = {IOP Publishing},
volume = {5},
number = {3},
pages = {035031},
author = {Birk, Joschka and Hallin, Anna and Kasieczka, Gregor},
title = {OmniJet-α: the first cross-task foundation model for particle physics},
journal = {Machine Learning: Science and Technology},
}

@article{Barman_2025,
   title={Large physics models: towards a collaborative approach with large language models and foundation models},
   volume={85},
   ISSN={1434-6052},
   url={http://dx.doi.org/10.1140/epjc/s10052-025-14707-8},
   DOI={10.1140/epjc/s10052-025-14707-8},
   number={9},
   journal={The European Physical Journal C},
   publisher={Springer Science and Business Media LLC},
   author={Barman, Kristian G. and Caron, Sascha and Sullivan, Emily and de Regt, Henk W. and de Austri, Roberto Ruiz and Boon, Mieke and Färber, Michael and Fröse, Stefan and Golling, Tobias and Lopez, Luis G. and Hasibi, Faegheh and Heinrich, Lukas and Ipp, Andreas and Kapoor, Rukshak and Kasieczka, Gregor and Kostić, Daniel and Krämer, Michael and Marco, Jesus and Otten, Sydney and Pawlowski, Pawel and Vischia, Pietro and Weber, Erik and Weniger, Christoph},
   year={2025},
   month=Sept }

@misc{he2026drsaiagenticairealworld,
      title={Dr.Sai: An agentic AI for real-world physics analysis at BESIII}, 
      author={Mingfeng He and Fayu Jiang and Junkun Jiao and Mingrun Li and Ke Li and Yipu Liao and Beijiang Liu and Tong Liu and Fazhi Qi and Zijie Shang and Weimin Song and Yue Sun and Xiongfei Wang and Hong Wang and Dongbo Xiong and Changzheng Yuan and Bolun Zhang and Zhengde Zhang and Xuliang Zhu},
      year={2026},
      eprint={2604.22541},
      archivePrefix={arXiv},
      primaryClass={hep-ex},
      url={https://arxiv.org/abs/2604.22541}, 
}

@misc{Aether,
    Author = "{Aether}",
    howpublished = {\url{https://aether.aiphys.cn}},
    note = {Accessed: 2026-05-03}
}

@article{hepscript,
    title = {HepScript: A Dual-Use DSL for Human-AI Collaborative Data Analysis Workflows in High-Energy Physics},
    Author = {Junkun Jiao and Tong Liu and Ke Li and Weimin Song and Yipu Liao and Bolun Zhang and Beijiang Liu and Chang-Zheng Yuan and Yue Sun},
    year={2026},
    eprint={2605.01423},
    archivePrefix={arXiv},
    primaryClass={hep-ex},
    url={https://arxiv.org/abs/2605.01423}, 
}

@article{opinionpaper,
    title = {AI agents, language, deep learning, and the next revolution in science},
    journal = {Frontiers of Physics},
    volume = {21},
    pages = {096401-},
    year = {2026},
    issn = {2095-0462},
    doi = {https://doi.org/10.15302/frontphys.2026.096401},
    howpublished = {\url{https://journal.hep.com.cn/fop/EN/10.15302/frontphys.2026.096401}},
    author = {Ke Li and Beijiang Liu and Bruce Mellado and Chang-Zheng Yuan and Zhengde Zhang},
}

@misc{SurveyA,
    Author = "{AI+HEP Community Survey (Part A)}",
    howpublished = {\url{https://docs.qq.com/form/page/DUWxTcFpjaVFzRUhD}},
    note = {Accessed: 2026-05-01}
}

@misc{SurveyB,
    Author = "{AI+HEP Community Survey (Part B)}",
    howpublished = {\url{https://docs.qq.com/form/page/DUU91U2NpcU9sU0Z6}},
    note = {Accessed: 2026-05-01}
}







\end{document}